\documentclass[letterpaper,american,english]{revtex4-1}
\usepackage[T1]{fontenc}
\usepackage[latin9]{inputenc}
\pdfoutput=1
\usepackage{amsmath}
\usepackage{amssymb}
\usepackage{graphicx}
\usepackage{esint}

\makeatletter

\pdfpageheight\paperheight
\pdfpagewidth\paperwidth

\usepackage{bbm}

\makeatother

\usepackage{babel}
\begin{document}

\preprint{BROWN-HET-1724}

\title{Black hole holography and mean field evolution}

\author{David A. Lowe}

\affiliation{Physics Department, Brown University, Providence, RI, 02912, USA}

\author{Larus Thorlacius}

\affiliation{University of Iceland, Science Institute, Dunhaga 3, IS-107, Reykjavik,
Iceland \foreignlanguage{american}{\\{\rm and}\\ }The Oskar Klein
Centre for Cosmoparticle Physics, Department of Physics, Stockholm
University, AlbaNova University Centre, 10691 Stockholm, Sweden}
\begin{abstract}
Holographic theories representing black holes are expected to exhibit
quantum chaos. We argue if the laws of quantum mechanics are expected
to hold for observers inside such black holes, then such holographic
theories must have a mean field approximation valid for typical black
hole states, and for timescales approaching the scrambling time. Using
simple spin models as examples, we examine the predictions of such
an approach for observers inside black holes, and more speculatively
inside cosmological horizons.
\end{abstract}
\maketitle

\section{Introduction}

Holographic theories offer a way to define string theory nonperturbatively
and address many of the outstanding questions in quantum gravity including
the black hole information paradox. In a holographic setting the dual
field theory provides a manifestly unitary description of gravitational
dynamics that in principle extends to the formation and subsequent
evaporation of black holes. The information about the microstate from
which a black hole is formed must then be preserved throughout the
evolution and released gradually to distant observers with the Hawking
radiation emitted over the lifetime of the black hole. 

In the present paper we address some key questions related to black
holes in holographic theories, in particular the holographic representation
of the black hole interior and near horizon region. Our main goal
is to holographically model observations made in a laboratory entering
a black hole in free fall and study deviations from bulk local quantum
field theory caused by nonlocal holographic interactions. This follows
up on and extends our previous work in \citep{Lowe:2015eba,Lowe:2016mhi}
on a toy model of black hole complementarity where local bulk evolution
for infalling degrees of freedom emerges from a highly nonlocal holographic
description via a mean field construction. The holographic degrees
of freedom describing the black hole region are represented by a discrete
collection of spin variables with long-range pairwise interactions
whose strength is adjusted so as to obtain finite energy states in
a large $N$ limit. With time the evolution of the mean field state
at a given spin site will diverge from the evolution generated by
the exact Hamiltonian of the spin system. This corresponds to the
breakdown of the local semiclassical bulk description of an infalling
observer inside a black hole due to holographic corrections and a
central result of \citep{Lowe:2015eba,Lowe:2016mhi} was that this
breakdown happens on the same timescale as the observer encounters
the curvature singularity.

Our strategy is to focus on the degrees of freedom that describe the
black hole interior and the near horizon region while eliminating
those that describe the asymptotic region far away from the black
hole. This is most economically represented by models with finite
dimensional Hilbert spaces, which can always be expressed in terms
of a finite number of interacting discrete spin variables. Our goal
is to then extract predictions about how observables depend on the
black hole microstate. The physics is governed by the interactions
between the spins which are \emph{a priori} unknown but should at
the very least be such that the model exhibits fast scrambling \citep{Sekino:2008he}
and the basic symmetries of the problem. In such models a version
of global thermalization occurs on a timescale of order 
\begin{equation}
t_{s}\sim\frac{1}{kT}\log S,\label{eq:scrambletime}
\end{equation}
where $T$ is the Hawking temperature of the black hole and $S$ is
the Bekenstein-Hawking entropy.

We note that our approach shares some common features with the Sachdev-Yi-Kitaev
(SYK) model \citep{sy_cite,kitaev}, which has received considerable
recent attention, but an important difference is that there one works
in a low temperature limit where a conformal symmetry arises. That
signals the model is describing an entire asymptotically anti-de Sitter
spacetime, and one must then be careful to disentangle the asymptotic
degrees of freedom from the black hole degrees of freedom in order
to address the kind of questions we are interested in here. In the
near horizon region and the interior of the black hole most of the
symmetries of the asymptotic spacetime are broken. One has at best
some remnant of rotational symmetry and an approximate time translation
symmetry, assuming the black hole evaporates slowly, which constrains
the possible interactions in the holographic Hamiltonian. 

The unitary evolution of a black hole in asymptotically flat spacetime
can be divided into four stages starting with its formation from matter
in a pure state undergoing gravitational collapse. The classical geometry
of the newly formed black hole will settle down fairly rapidly on
a timescale of order the black hole mass in natural units into a quasi-stationary
state, which for an uncharged and non-rotating black hole will be
well described locally as a Schwarzschild geometry. This is followed
by two stages of slow evaporation due to the emission of Hawking radiation.
The two stages are distinguished by the nature of the quantum entanglement
among the degrees of freedom that make up the black hole and outgoing
Hawking radiation. During the earlier stage, often referred to as
a young black hole, the majority of these degrees of freedom reside
inside or near the remaining black hole and the entanglement entropy
of the subsystem consisting of the Hawking radiation that has escaped
from the near region is then very close to its maximal possible value.
After the so-called Page time \citep{Page:1993df}, when the area
of the remaining black hole has been reduced to half its initial value,
the roles are reversed and the Hawking radiation carries more than
half the degrees of freedom. The state of the subsystem consisting
of an old black hole, past its Page time, and its near environment
is then maximally entangled with the long train of Hawking radiation
that has already been emitted. The fourth stage comes at the end of
the black hole lifetime when the Hawking temperature approaches the
Planck temperature and no known gravitational theory can provide a
reliable description of the physics. This is not a major concern however
as the violent final stage is very short lived and the black hole
has already given up almost all its original quantum information before
entering into it. In this work we will only be concerned with questions
involving the two intermediate stages of slow evaporation but it will
be important to distinguish between young and old black holes and
their different entanglement structure in the holographic description
of the interior geometry. 

Hawking emission gradually transfers degrees of freedom from the black
hole to outgoing radiation propagating in the far region and the black
hole should strictly speaking be viewed as an open quantum system
whose degrees of freedom are depleted with time. We can sidestep this
complication since the characteristic timescale of the dynamics we
are exploring is the scrambling time \eqref{eq:scrambletime} and
this is parametrically short compared to the black hole lifetime.
Thus the number of Hawking quanta emitted in a scrambling time is
very small compared to the total number of qubits required to describe
the black hole. For our purposes here, it will thus be sufficient
to consider a spin model with a fixed large number $N$ of spins,
where $N$ is proportional to the Bekenstein-Hawking entropy of the
remaining black hole. A young black hole can then be approximately
represented by a pure state where the spins are maximally entangled,
ignoring any error made by leaving out the degrees of freedom of the
emitted Hawking radiation. In an old black hole on the other hand
the spins encoding the black hole interior are entangled with the
previously emitted Hawking radiation and are described by a maximally
mixed spin state. 

A class of spin models with highly nonlocal interactions was studied
in \citep{Lashkari:2011yi} and it was shown that the resulting dynamics
exhibits a form of fast scrambling for the states of interest. These
models have $N$ discrete spins $\vec{\sigma}_{i}$ with pairwise
interactions 
\begin{equation}
H=\sum H_{ij}\,,\label{eq:spin_hamiltonian}
\end{equation}
where the sum ranges over unordered pairs of sites and the interactions
are dense in the sense that the number of pairs with non-vanishing
$H_{ij}$ scales as $N^{2}$. The interactions therefore have infinite
range and the dynamics is insensitive to the dimension of the embedding
spacetime. In order to have a sensible energy per degree of freedom
in the large $N$ limit the strength of each two-body interaction
must be bounded $||H_{ij}||<c/N$, where the norm $||\mathcal{O}||$
of an operator is given by the absolute value of its largest eigenvalue
and $c$ is a constant.

Spin Hamiltonians of this type were further studied in \citep{Lowe:2016mhi}
as simple holographic models of stretched horizon dynamics in the
context of black hole complementarity. The local Hamiltonian dynamics
in a radially infalling frame was modeled by a mean field Hamiltonian
that depended on the initial state of the spin system. When the initial
state is a product of two factors, one describing the black hole and
the other representing a small ``laboratory'' entering the black
hole in free fall, the decoherence between the local mean field evolution
of the laboratory state and the exact holographic evolution takes
place on timescales comparable to the scrambling time. This suggests
that strong quantum gravity effects only occur well after the laboratory
enters the black hole and not immediately upon crossing the horizon.
The arguments presented in \citep{Lowe:2016mhi} apply to a class
of spin models which is hopefully sufficiently generic to capture
some of the behavior of actual holographic systems. The description
is very crude however and to correctly describe the short-range Newtonian
gravity limit of freely falling laboratories in detail would require
a much more detailed specification of the holographic model, which
for now we steer clear of. 

In gauge/gravity duality in general, the map between the holographic
model and the gravitational spacetime is extremely nonlocal. In the
present case, we are already led to use a maximally nonlocal interaction
between the spins to generate fast scrambling. We may view the spins
as living on a lattice on a spherical surface embedded in the black
hole spacetime. For example, we can consider approximately spherically
symmetric states in four-dimensional spacetime, in which case the
lattice will span the $\theta,\phi$ coordinates of a Schwarzschild
solution.

The observables we will be interested in correspond to experiments
conducted in freely falling laboratories, which we may then represent
as a site on the lattice where the state is of the form
\begin{equation}
|\psi\rangle=|\psi_{lab}\rangle\otimes|\psi_{bh}\rangle\label{eq:labstate}
\end{equation}
where $|\psi_{lab}\rangle$ is a pure state of a spin on a lattice
site. Here the term ``spin'' is used to represent any finite-dimensional
Hilbert subspace, which encodes the full quantum state of the laboratory,
and its onsite Hamiltonian can be arbitrary. For a young black hole
$|\psi_{bh}\rangle$ is a pure maximally entangled state on the remaining
lattice sites. An old black hole on the other hand is maximally entangled
with previously emitted Hawking radiation whose degrees of freedom
are not included in the spin model and in this case the state $|\psi_{bh}\rangle$
should be replaced by a density matrix describing a maximally mixed
state. 

Scrambling is accomplished via the nonlocal coupling between different
sites. Time evolution with respect to the holographic Hamiltonian
will evolve the state \eqref{eq:labstate} forward in time. The main
problem in mapping holographic time evolution to bulk time evolution
involves correctly identifying the relevant bulk timeslices. In this
regard, we follow \citep{Lowe:2015eba} where a rather generic freely
falling Planck lattice was considered as a bulk regulator. In \citep{Lowe:2016mhi}
we found that many of the properties of holographic states are reproduced
when matched with such a bulk description. On the bulk gravity side
this evolution corresponds to propagating the laboratory inwards along
a radial timelike geodesic. Since the holographic Hamiltonian eventually
evolves the state toward a stationary thermal state, we are able to
match the holographic time coordinate $t$ with the bulk Killing vector
time, and we assume this matching can be carried out near the black
hole a few Schwarzschild radii outside the horizon. 

If we form the reduced density matrix of the laboratory site, by tracing
over the Hilbert subspace associated with all other sites, the evolution
will behave as an open quantum system \citep{Davies:1976nb,Benatti:2009:DIC:1610379}.
In particular, the state $|\psi_{lab}\rangle$ will not experience
local unitary evolution due to interactions with other sites. Ordinary
local interactions with the surrounding spacetime account for part
of this effect, but the dominant effect comes down to the nonlocal
interaction with distant sites. One of the main questions we are interested
in is to quantify the degree to which these nonlocal effects disrupt
the experience of unitary quantum mechanics for the falling laboratory. 

To study this we formulate a mean field approximation for the evolution
of the state \eqref{eq:labstate}. Time dependent mean field approximations
have previously been studied in interacting fermion systems in \citep{BARDOS2003665,Benedikter2014}.
Such a mean field Hamiltonian by construction will be sitewise local,
and Hermitian, and will be guaranteed to give a unitary evolution
of the pure lab state $|\psi_{lab}\rangle$. However the choice of
mean field Hamiltonian depends on the state $|\psi_{bh}\rangle$ so
that the dynamics overall will not satisfy the superposition principle.
By comparing the trace distance between the time evolution with respect
to the exact Hamiltonian and the mean field Hamiltonian, one finds
\begin{equation}
\left\Vert \rho_{lab}(t)-\rho_{lab}^{MF}(t)\right\Vert _{1}<\frac{1}{S_{bh}}e^{kTt}\label{eq:tracedist}
\end{equation}
for a general class of spin models (matching $c=kT$ in the notation
of \citep{Lowe:2016mhi}). We conclude that as long as $t<t_{s}$,
with the scrambling time $t_{s}$ given by \eqref{eq:scrambletime},
then the nonlinear quantum mechanics violations of mean field are
under control, and likewise the non-unitary effects of the exact evolution
may be ignored. Thus for a finite time interval, the freely falling
laboratory experiences the usual rules of quantum mechanics governed
by the mean field Hamiltonian.

As we discuss more in the following, one may take a variety of different
mean field Hamiltonia which are, in part, in correspondence with different
choices of timeslicings of the bulk geometry. In previous work, we
emphasized that there is a clear correspondence with slicings corresponding
to freely falling coordinates (and close deformations thereof) where
the timescale \eqref{eq:scrambletime} arises as an upper bound on
the time at which the laboratory hits the curvature singularity of
the black hole geometry. Thus there seems to be a correspondence between
singularity approach in the spacetime geometry and the breakdown of
the mean field approximation in the holographic theory.

In the present paper our main goal is to study the rather detailed
predictions that emerge for the corrections to local quantum field
theory a freely falling laboratory will experience in the vicinity
of a spacetime horizon. The clearest prediction is a shift of physical
couplings in the laboratory by an amount that depends on the state
of the black hole. We compute this in a number of examples where we
find
\begin{equation}
\delta H_{ii}=\mathrm{Tr}_{i^{c}}\rho_{MF}(t)\sum_{j\neq i}H_{ij}\sim\frac{kT_{bh}}{S_{bh}^{1/2}}\,,\label{eq:couplingshift}
\end{equation}
Evaluating this for a four-dimensional Schwarzschild black hole the
couplings shift by order $1/M^{2}$ producing a timescale for observation
of order $M^{2}$. This is significantly longer than the scrambling
time, so will be difficult to detect by an interior observer \citep{Lowe:2015eba}.
On the other hand, this may lead to interesting effects for observers
who remain outside the horizon, or observers in expanding universes
who are potentially very long-lived. 

Because the shift in the couplings depends on the microstate of the
black hole, this may be viewed as a concrete realization of the soft
hair proposal of \citep{Hawking:2016msc,Hawking:2016sgy}. To properly
complete that idea, some regularization is needed to thin the infinite
soft hair out to something that might account for the finite black
hole entropy. The holographic construction in the present paper gives
a nice example of such a regularization.

We also study the spectrum of maximally entangled states in the holographic
theory. By construction, we are always free to add a constant to the
Hamiltonian to match the overall energy with the bulk energy. More
interesting is the distribution of states in the theory. With a suitable
choice of model parameters we find the width of the energy spectrum
to be $\sim kT$, corresponding to a highly degenerate black hole
system with a spectral width of order the energy of a single Hawking
particle. 

We close with some thoughts on crossover models which will be a necessary
generalization of these models if the goal is to generate both fast
scrambling and the correct short-distance dynamics in the bulk. Broadly
speaking one expects a spin-spin correlator to fall off on a length/timescale
of order $1/kT$ in a model with the correct short distance behavior.
Such behavior can be generated by including both infinite range and
nearest neighbor spin-spin interactions, and interesting critical
points can be obtained by tuning the respective couplings. We hope
to return to the consideration of such models in future work.

\section{\label{sec:Scrambling-in-spin}Scrambling in spin models}

Quantum chaos in the usual sense is believed to appear in models with
higher order interactions \citep{kitaev}. In the present work we
will use a much simpler model based on pairwise interactions, and
define our notion of scrambling using the trace distance between states,
as in \eqref{eq:tracedist}. Note that not all states will scramble
rapidly in the models considered here. Moreover in the case of integrable
spin models, there are an infinite number of conserved quantities,
so many observables will not scramble at all. For the case at hand,
we will be interested in a particular class of pure states that are
maximally entangled, which exhibit fast scrambling even with simple
integrable Hamiltonia. The main property we will need is simply the
``infinite''-range interactions, coupling each site to every other. 

In future, we hope to return to this set of questions in more complex
models such as SYK, which in certain limits appear to saturate the
scrambling time (at least when defined in terms of out-of-time four-point
functions in a thermal state). The hope would be that in more general
models generic initial states would scramble rapidly, in line with
the conjectured behavior for the dual black holes. For the present,
by restricting to initial states that are close to maximal entanglement,
we can sidestep this issue, and still explore whether small laboratory
subsystems still experience a breakdown of quantum mechanics with
respect to their local mean field Hamiltonians.

In earlier work, Lieb-Robinson bounds have been used to derive bounds
on the trace distance \citep{Lashkari:2011yi,Lowe:2014vfa,Lowe:2016mhi}.
The advantage of this approach is that it works for any choice of
pairwise Hamiltonian (with arbitrary ``local'' onsite couplings)
and any choice of state. However one obtains only upper bounds on
the behavior of the trace distance. Our first task in the present
work is to see to what extent these bounds are saturated. We will
find via direct numerical calculation that for sufficiently early
times a bound of the form \eqref{eq:tracedist} is saturated. As the
reduced density matrix approaches maximal mixing, the growth levels
off, and exhibits Loschmidt echo phenomena \citep{Peres,PhysRevLett.86.2490,Zunkovic20150160}.
In a more detailed model of quantum chaos, we expect this echo regime
to be replaced by approximately constant trace distance indicative
of a maximally mixed state, with recurrence happening over an inaccessible
double exponential timescale \citep{PhysRev.107.337,Page:1994dx}.
However for the purposes of the present paper, where we focus on interior
operators, it is sufficient to study the operators in the early time
regime, only involving evolution approaching that of the scrambling
time \eqref{eq:scrambletime}.

\subsection{Modeling a young black hole}

Having described the general setup, let us proceed to computation.
In the following we will use the infinite-range Heisenberg model 
\begin{equation}
H=-\frac{J}{N}\sum_{i,j=1}^{N}\vec{s}_{i}\cdot\vec{s}_{j}\,,\label{eq:heisenberg1}
\end{equation}
where a spin $1/2$ degree of freedom on each site interacts with
the rest. This is a particularly simple representative of the class
of spin models in \eqref{eq:spin_hamiltonian}, with a uniform pairwise
interaction and a single coupling constant $J$. Despite its simplicity
the infinite-range Heisenberg model satisfies a basic requirement
for scrambling that a local spin $\vec{s}_{i}$ does not commute with
the Hamiltonian. For our initial state we take a state of the form
\eqref{eq:labstate} where $|\psi_{bh}\rangle$ is chosen to be maximally
entangled. A maximally entangled state may be constructed by taking
a pairwise product of singlet states
\begin{equation}
|\psi_{max}\rangle=\bigotimes_{i=i}^{(N-2)/2}\frac{1}{\sqrt{2}}\left(\left|\uparrow\right\rangle _{2i+1}\left|\downarrow\right\rangle _{2i+2}-\left|\downarrow\right\rangle _{2i+1}\left|\uparrow\right\rangle _{2i+2}\right)\label{eq:spinsinglets}
\end{equation}
Now one might be concerned this is a special choice of state. In general
therefore we select a random unitary transformation $U_{ran}$ on
the $2^{N-2}$ dimensional Hilbert space of the black hole sector,
sampled with the natural unitary measure, and construct
\[
|\psi_{bh}\rangle=U_{ran}|\psi_{max}\rangle
\]
The state $|\psi_{bh}\rangle$ remains close to being maximally entangled,
and may then be used to compute \eqref{eq:tracedist} for generic
maximally entangled states. 

\begin{figure}

\includegraphics{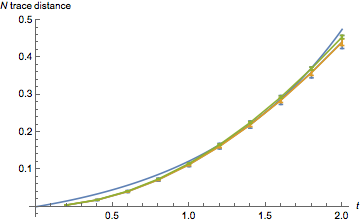}\caption{\label{fig:puretrace}$N\left\Vert \rho_{lab}(t)-\rho_{lab}^{MF}(t)\right\Vert _{1}$
is plotted for various values of $N\leq14$. The error bars indicate
an uncertainty due to picking different initial states. The numerical
data is bounded by the curve $0.065\left(\exp\left(1.06t\right)-1\right)$
shown in blue in the plot.}

\end{figure}

In figure \ref{fig:puretrace} the trace distance $N\left\Vert \rho_{lab}(t)-\rho_{lab}^{MF}(t)\right\Vert _{1}$
is computed and averaged over random unitary matrices for values of
$N\leq14$. The early time behavior shows a time dependence that converges
as $N$ increases to an $N$-indepedent form. Likewise the error bars
associated with the average over initial states are small. Finally
the function is bounded by a function of the form $0.065\left(\exp\left(1.06t\right)-1\right)$,
allowing us to extract the scrambling time $t\sim\beta\log N$ as
a characteristic timescale.

\begin{figure}

\includegraphics{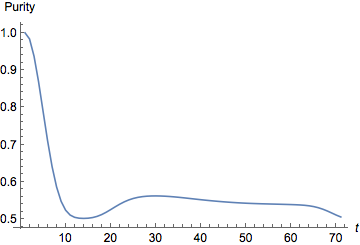}\caption{\label{fig:purity}The purity $\mathrm{Tr}\rho_{lab}^{2}(t)$ decreases
from 1 to close to maximal mixing. Then there is a partial recurrence,
known as Loschmidt echo.}

\end{figure}

\begin{figure}

\includegraphics{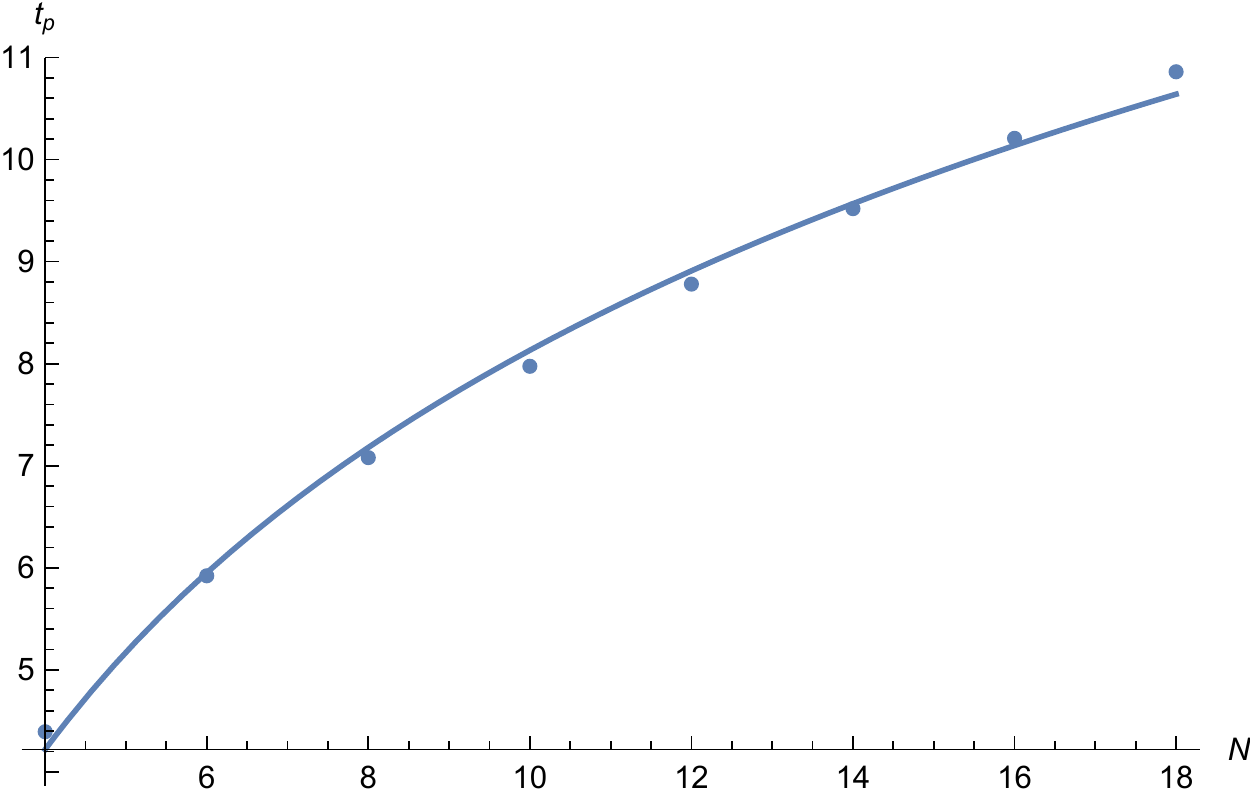}\caption{\label{fig:mixtime}The mixing time defined by the minimum of the
purity is plotted for different values of $N$. The blue line shows
a fit to $t_{p}=-1.70+4.26\log N$.}

\end{figure}

On somewhat longer timescales than shown in figure \ref{fig:puretrace}
the purity $\mathrm{Tr}\rho_{lab}^{2}(t)$, which starts at $1$,
approaches that of a maximally mixed state (or $1/2$ for a 2-state
system). After this a Loschmidt echo develops, as shown in figure
\ref{fig:purity}. If we identify the mixing timescale with the minimum
of the purity, this is well fit by a logarithmic function $t_{p}=-1.70+4.26\log N$
as shown in figure \ref{fig:mixtime}, providing another way to see
the scrambling time emerge from the evolution.

\subsection{Modeling an old black hole}

As a check on the previous results it is interesting to instead begin
with the black hole sector in a maximally mixed density matrix. With
the present computations, this is most easily accomplished by purifying
the state. One way to accomplish this is to begin with the initial
state \eqref{eq:spinsinglets} but set the Hamiltonian to vanish on
sites indexed by $i$ even, and unchanged when $i$ is odd. The even
sites then undergo trivial time evolution, but remain maximally entangled
with the odd sites. Finally we may act with a random unitary transformation
on the Hilbert subspace of the odd sites to generate a typical entangled
state.

\begin{figure}

\includegraphics{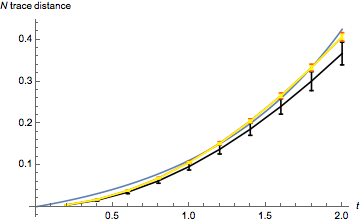}\caption{\label{fig:mixedtrace}$N\left\Vert \rho_{lab}(t)-\rho_{lab}^{MF}(t)\right\Vert _{1}$
is plotted for maximally mixed black hole density matrices, representing
an old black hole. The behavior is very similar to figure \ref{fig:puretrace}.}

\end{figure}

The results for $N_{tot}\leq14$ (\emph{i.e.} $N_{bh}\leq7$) are
shown in figure \ref{fig:mixedtrace}. Up to numerical errors the
results are the same as in the previous subsection for the early time
behavior. Again this shows that $N\left\Vert \rho_{lab}(t)-\rho_{lab}^{MF}(t)\right\Vert _{1}$
approaches an $N$ independent limit, that is bounded by a function
of the form $0.061\left(\exp\left(1.04t\right)-1\right)$.

\subsection{Summary}

We have provided evidence that the early time behavior of the trace
distance takes a universal form as $N$ becomes large. This comes
close to saturating a bound of the form $\frac{a}{N}(exp\left(bt\right)-1)$
where $a,b$ are $N$ independent constants. The scrambling time $t\sim\log N$
then naturally emerges from this construction.

\section{Mean field theory}

\subsection{Maximally entangled states in spin models}

A maximally entangled state is unitary equivalent to a product state
of the form \eqref{eq:spinsinglets}, where for simplicity we imagine
each site contains a spin $1/2$ degree of freedom. As shown in \citep{Lowe:2016mhi},
we may start with a generic maximally entangled state and perform
a unitary transformation to get such a state. Such a unitary transformation
does not necessarily commute with the Hamiltonian.

Now if we coarse grain, combining the sites $k$ and $k+1$ into a
single site, and the state on each pair into a pure state on the coarse-grained
lattice, we are left with an unentangled pure state
\[
|\psi\rangle=\prod_{k=2i}|\chi\rangle_{k}
\]
Our goal in this section is to study observables in this class of
states in the coarse grained spin model, where the generic state is
such an unentangled state. As will become clear below, it is sufficient
to consider the spin model in a high temperature limit to compute
the observables of interest. We focus our attention on the infinite-range
Heisenberg model as a representative of the more general class of
nonlocal spin models \eqref{eq:spin_hamiltonian}. In the previous
section we presented numerical evidence that even in this simple model
the nonlocal spin dynamics fast scrambles maximally entangled states.
In the following we carry out a thermodynamic analysis of the energy
spectrum and argue that with the appropriate choice of parameters
it matches the spectrum of an evaporating black hole.

\subsection{Saddle point analysis of the infinite-range Heisenberg model}

Consider $N$ spins interacting via the Hamiltonian \eqref{eq:heisenberg1}
where the spin operator $\vec{s_{i}}$ generates an $SU(2)$ rotation
on a two-component qubit at site $i$ and $J>0$ is a constant that
will be adjusted later on to match the behavior of black hole states.
For convenience we have included diagonal terms with $i=j$ in the
double sum in the Hamiltonian. These terms contribute a constant to
the total energy and by including them the Hamiltonian is simply the
square of the total spin,
\[
H=-\frac{J}{N}\left(\sum_{i=1}^{N}\vec{s}_{i}\right)^{2}\,.
\]
The energy eigenvalues are given by $E_{S}=-JS(S+1)/N$ where the
total spin quantum number $S$ is integer spaced and ranges from $N/2$
down to $0$ or $1/2$ depending on whether $N$ is even or odd. 

The finite temperature behavior of the system is encoded in the partition
sum,
\[
Z[\beta]=\mathrm{Tr}\,[e^{-\beta H}]=\sum_{S}d(S,N)\,e^{\beta JS(S+1)/N},
\]
where $d(S,N)$ is the number of states with total spin $S$ appearing
in the $N$-fold tensor product of fundamental representations \citep{Aste:2015lla},
\[
d(S,N)=\frac{(2S+1)^{2}}{N+1}\left(\begin{array}{c}
N+1\\
\frac{N}{2}-S
\end{array}\right).
\]
At a fixed large value of $N$ the number of states has a maximum
at $S\simeq\sqrt{N/2}$ and the partition sum is dominated by terms
where both arguments of the binomial coefficient are large. In this
case we can use Stirling's approximation, 
\[
\left(\begin{array}{c}
N+1\\
\frac{N}{2}(1-x)
\end{array}\right)\simeq\frac{2^{N+\frac{3}{2}}}{\sqrt{\pi N}}(1-x)^{-1/2}(1+x)^{-3/2}e^{-\frac{N}{2}\left[(1-x)\log(1-x)+(1+x)\log(1+x)\right]},
\]
where $x=2S/N$. The sum over $S$ can then be approximated by an
integral over $x$, 
\[
Z\left[\beta\right]=\frac{2^{N+\frac{1}{2}}N^{\frac{3}{2}}}{\sqrt{\pi}}\intop_{0}^{1}dx\,f(x)e^{Ng(x)},
\]
with
\[
f(x)=\frac{\left(x+\frac{1}{N}\right)^{2}e^{\beta Jx/2}}{\left(1-x\right)^{1/2}\left(1+x\right)^{3/2}},
\]
and 
\[
g(x)=\frac{1}{4}\beta Jx^{2}-\frac{1}{2}\left(1-x\right)\log\left(1-x\right)-\frac{1}{2}\left(1+x\right)\log\left(1+x\right).
\]
The integral can be evaluated in a saddle point approximation valid
at large $N$. The resulting saddle point equation, 
\[
x=\tanh\left(\frac{1}{2}\beta Jx\right),
\]
reveals a second order phase transition at a critical temperature
$kT_{c}=J/2$. 

\begin{figure}
\includegraphics{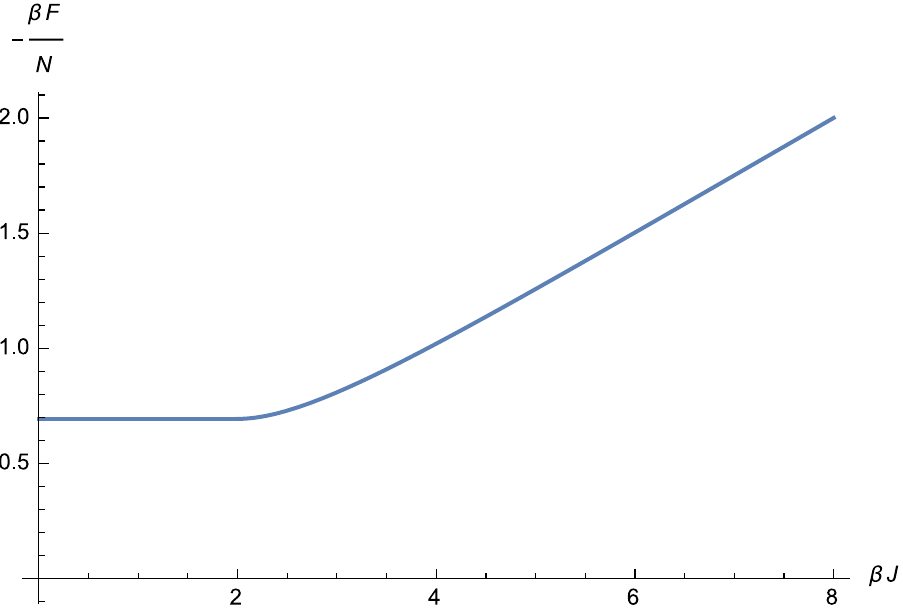}

\caption{\label{fig:Free-energy-per}Free energy per site at large $N$ in
the infinite range Heisenberg model.}
\end{figure}

In the low temperature phase $\beta J>2$ the saddle point equation
has a non-trivial solution at $x=x_{*}>0$ and the free energy is
given by
\[
-\beta F=\log Z=N(\log2+g(x_{*}))+O(\log N).
\]
Figure \ref{fig:Free-energy-per} plots the saddle point value of
the free energy per site, as a function of the dimensionless variable
$\beta J$ in the limit of large $N$.

It is, however, the high temperature limit $\beta J\ll1$ that is
of interest for modeling black hole behavior. In the high temperature
phase the saddle point sits at the endpoint of the integration range
$x=0$ and in this case the leading large $N$ contribution to the
free energy is simply
\[
-\beta F=\log Z=N\log2+\cdots.
\]
The energy vanishes at leading order, 
\[
U=-\frac{\partial}{\partial\beta}\log Z=0+\cdots,
\]
and the entropy is given by
\[
S=k\left(1-\beta\frac{\partial}{\partial\beta}\right)\log Z=kN\log2+\cdots.
\]
We need to include subleading $N^{0}$ terms in our saddle point analysis
in order to see how the thermodynamic variables depend on temperature.
A straightforward calculation gives
\begin{align*}
-\beta F & =N\log2-\frac{3}{2}\log\left(1-\frac{\beta J}{2}\right),\\
S & =kN\log2-\frac{3k}{2}\log\left(1-\frac{\beta J}{2}\right)-\frac{3k\beta J}{4-2\beta J},\\
U & =-\frac{3J}{4-2\beta J}.
\end{align*}
The negative sign energy is consistent with $H$ in equation \eqref{eq:heisenberg1}
being negative definite. We note that the entropy scales linearly
with the number of sites in the high temperature limit $\beta J\rightarrow0$
while the width of the energy spectrum remains finite at large $N$,
\[
\left(\Delta U\right)^{2}=\frac{\partial^{2}}{\partial\beta^{2}}\log Z=\frac{3J^{2}}{2\left(2-\beta J\right)^{2}}.
\]

\subsection{Matching black hole parameters}

If we shift $U$ by a constant so as to match the black hole energy
($U=M)$ and dial the coupling so that 
\begin{equation}
J\sim kT_{bh}\,,\label{eq:heisenbergcoupling}
\end{equation}
corresponding to matching the scrambling time in the model \citep{Lowe:2016mhi}
to \eqref{eq:scrambletime}, then the high temperature phase of this
simple spin model exhibits precisely the sort of highly degenerate
energy spectrum that one expects for an evaporating black hole. In
particular the width of the black hole states, which is independent
of the constant shift in $U$, matches that of a Hawking quantum.

The self consistency of considering the high temperature phase ($\beta J\ll1$)
of the spin model can be checked as follows. The entropy scales linearly
with the number of sites and is also proportional to the black hole
area, $N\sim S\sim r_{s}^{D-2}$, where $r_{s}$ is the Schwarzschild
radius and D is the spacetime dimension. The black hole temperature
scales as $T_{bh}\sim r_{s}^{-1}\sim N^{-1/(D-2)}$ and thus the identification
\eqref{eq:heisenbergcoupling} ensures that $\beta J\rightarrow0$
in the large $N$ limit for any finite value of $\beta$. In other
words the spin model is effectively in a high temperature limit for
the parameter values that match the black hole physics we wish to
model. 

The above results may also be used to infer the mean field shift in
the onsite Hamiltonian,
\begin{equation}
(\Delta H_{ii})^{2}=\frac{1}{N}(\Delta U)^{2}\sim\frac{\left(kT_{bh}\right)^{2}}{S_{bh}}.\label{eq:shift}
\end{equation}
The onsite Hamiltonian is to be viewed as the worldline Hamiltonian
for a free falling observer in the bulk spacetime. Therefore this
shift \eqref{eq:shift} gives an estimate of the size of the potentially
observable effect a local observer can detect that arises from the
nonlocal holographic physics of the exact evolution. For black holes,
this kind of energy shift will be hard to measure in free fall since
the lifetime of such an infalling observer with respect to the holographic
time will be bounded by \eqref{eq:scrambletime} by the arguments
of \citep{Lowe:2015eba}. However an exterior observer who is long-lived,
and able to conduct experiments on the stretched horizon should be
able to avoid such a restriction. In this case, such an observer can
potentially distinguish the microstates of the black hole. One can
perhaps view this dependence of local observables on the black hole
microstates as a concrete realization of the soft hair proposal of
\citep{Hawking:2016msc}.

It is also interesting to apply the above kind of reasoning to a de
Sitter horizon. Let us for the sake of argument assume that a similar
holographic description involving a finite dimensional spin model
can be developed for the degrees of freedom associated with a stretched
de Sitter horizon. In this case a freely falling observer inside a
Hubble volume can in principle live long enough to make the observation
of energy shifts analogous to \eqref{eq:shift} observable, allowing
a de Sitter observer to gain information about the horizon microstate.
The analogy with the black hole problem leads to a more worrying concern
that local quantum mechanics will break down for such an observer
after a de Sitter scrambling time $t_{s}\sim L_{dS}\log\left(L_{dS}\right)$.
The construction of a convincing holographic description of de Sitter
spacetime remains an open problem. We hope to return to this issue
in future work.
\begin{acknowledgments}
This research was supported in part by DOE grant de-sc0010010, Icelandic
Research Fund grant 163422-051, the University of Iceland Research
Fund, and the Swedish Research Council under contract 621-2014-5838.
The authors also wish to thank the Simons Center for Geometry and
Physics, Stony Brook University, for hospitality and support during
the final stages of the project. 
\end{acknowledgments}

\bibliographystyle{utphys}
\bibliography{mean_field}

\end{document}